\documentclass[twocolumn,
pre,showpacs,floatfix,amsmath,amsfonts]{revtex4}
\usepackage[latin9]{inputenc}
\usepackage{color}
\usepackage{amsmath}
\usepackage{amssymb}
\usepackage{graphicx}
\usepackage{esint}

\makeatletter
\@ifundefined{textcolor}{}
{%
 \definecolor{BLACK}{gray}{0}
 \definecolor{WHITE}{gray}{1}
 \definecolor{RED}{rgb}{1,0,0}
 \definecolor{GREEN}{rgb}{0,1,0}
 \definecolor{BLUE}{rgb}{0,0,1}
 \definecolor{CYAN}{cmyk}{1,0,0,0}
 \definecolor{MAGENTA}{cmyk}{0,1,0,0}
 \definecolor{YELLOW}{cmyk}{0,0,1,0}
}

\usepackage{color}
\usepackage{graphics}
\usepackage{epsfig}
\newcommand{\be}{\begin{equation}}
\newcommand{\ee}{\end{equation}}
\newcommand{\bea}{\begin{eqnarray}}
\newcommand{\eea}{\end{eqnarray}}
\newcommand{\bes}{\begin{subequations}}
\newcommand{\ees}{\end{subequations}}

\newcommand{\sech}{\mathrm{sech}}
\newcommand{\acosh}{\mathrm{acosh}}
\newcommand{\bU}{\mathbf{U}}

\makeatother

\begin{document}

\title{Asymmetric Perfect Absorption and Lasing of Nonlinear Waves by a Complex $\delta$-potential}

 \author{Dmitry A. Zezyulin$^{1}$   and Vladimir V. Konotop$^{2,3}$}

\affiliation{  $^{1}$~ITMO University, St. Petersburg 197101, Russia\\
	$^{2}$~Centro de F\'{\i}sica Te\'orica e Computacional, Faculdade de Ci\^{e}ncias da Universidade de Lisboa, Universidade de Lisboa, Campo
	Grande, Edif\'{\i}cio C8, Lisboa 1749-016, Portugal\\
	$^{3}$~Departamento de F\'{\i}sica, Faculdade de Ci\^{e}ncias da Universidade de Lisboa, Campo Grande, Edif\'{\i}cio C8, Lisboa 1749-016, 
	Portugal}

\begin{abstract}
Spectral singularities and the coherent perfect absorption are two interrelated concepts that  have originally   been introduced and studied for  linear waves interacting with complex potentials. In the meantime,  {the} distinctive asymptotic behavior of perfectly absorbed waves suggests  {considering} possible generalizations of these phenomena for nonlinear waves. Here we address perfect absorption of  nonlinear waves by an idealized infinitely narrow dissipative potential  {modeled} by a Dirac $\delta$-function with an imaginary amplitude. Our main result is the existence of perfectly absorbed flows whose spatial amplitude distributions are asymmetric with respect to the position of the absorber. These asymmetric states do not have a linear counterpart. Their linear stability is verified numerically. The nonlinear waveguide also supports symmetric and constant-amplitude perfectly absorbed flows.  {The} stability of solutions of the latter type can be confirmed analytically.
\end{abstract}


\maketitle
\section{Introduction}

The concept of spectral singularities (SSs) is known in mathematics already for  {a}  long time~\cite{Naimark,Schwartz}. The related physical phenomenon, known today as coherent perfect absorption (CPA)~\cite{Stone}, was discovered independently~\cite{CPA1,CPA2,CPA3} {(see \cite{CPA4} for a chronological review on the topic)} and is characterized by the asymptotic behavior of the field corresponding to only incoming wave. The link between mathematical singularities and asymptotic behaviour of the respective solutions is established by the theorem due to  Vainberg \cite{Vainberg}. In the last decade, the interest in physical effects related to the  SSs was revitalized due do to a  series of   works~\cite{Mstafa,Longhi,Stone} {establishing direct links between mathematical properties of SSs and their relevance for physical applications, as well as due to the first experimental implementation of a CPA}~\cite{Science}. 

{The early proposals of CPA addressed  {the} absorption of the electromagnetic radiation by a layer with the complex-valued dielectric permittivity. This, in particular, can be implemented using a confined plasma layer \cite{CPA2}. More recently, CPA was reported for  a variety of photonic systems, including plasmonic metasurfaces, graphene-based systems  electromagnetic waves  interacting with graphene and  plasmonic metasurfaces, microcavities, etc., see a recent review \cite{review} on  physical applications of photonic coherent perfect absorbers. Moreover, the paradigm of CPA was enriched by addressing  {the} absorption of waves of various nature, such as acoustic waves interacting with a fluid absorber \cite{acoustic} and quantum superfluids  depleted by a  focused  electron beam applied to an atomic Bose-Einstein condensate (BEC) \cite{Mullers2018}.}

By its definition, SS  is an essentially linear object. However, considering it from the physical point of view, i.e., focusing on the  distinctive asymptotic behavior of the solutions associated with SSs, one can extend the paradigm to nonlinear setups. One of the ways to do this is  {by} using nonlinear properties of a confined absorbing layer embedded in a linear medium~\cite{nonlin-pot1,nonlin-pot2}. Another possible way is to consider waves propagating in a nonlinear medium and interacting with a linear absorbing potential. This generalization of the concept of CPA was suggested in \cite{ZezKon2016} and validated in experiments with atomic BECs  \cite{Mullers2018}. Furthermore,  {the two types of  nonlinearities} (that of the potential and that of the medium) can be combined~\cite{ZOK}. Strictly speaking, in the nonlinear case there is no interference of states, and thus the coherence loses the meaning it has in the linear theory, but one still can consider perfect absorption. 

These recent developments raise new questions in the theory of perfect absorption, such as the effects of eventual instabilities and nonlinearity-induced symmetry-breaking. The aim of this paper is to show that   waveguides with a spatially uniform nonlinearity and   a localized dissipation in the form of the idealized $\delta$-function support perfectly absorbed modes with asymmetric amplitude distribution. In sharp contrast to constant-amplitude CPA solutions, nonlinear asymmetric currents cannot be reduced to the linear limit by decreasing the background  solution intensity.  

In a more general context of   the nonlinear waves theory, the found asymmetric states are remarkable, because they are supported by only  a single absorbing layer. This situation contrasts with the well-studied  conventional spontaneous  symmetry-breaking of nonlinear waves which typically  requires  a double-well spatial potential (see e.g. \cite{symm-break1,symm-break2,symm-break3,symm-break4}) or results from the competition between spatially inhomogeneous  linear and nonlinear potentials \cite{symm-break5} .

The rest of our paper is organized as follows. In {Section}~\ref{sec:model}  we introduce  the main model and provide some preliminary discussion,  and in {Section}~\ref{sec:results} we present and discuss the main results of the study. Section~\ref{sec:concl} concludes the paper.

\section{The model}
\label{sec:model}

We  consider the spatially one-dimensional   defocusing nonlinear Schr\"odinger equation  (NLSE)
\begin{eqnarray}
\label{eq:main}
i\Psi_t  = -\Psi_{xx}  - i\gamma \delta(x)\Psi + |\Psi|^2\Psi,
\end{eqnarray}
where $\gamma\ne 0$ is a real parameter which governs the strength of the dissipation (for $\gamma>0$) or energy gain (for $\gamma<0$), and $\delta(x)$ is the Dirac delta function. {The model (\ref{eq:main}) was introduced in Ref.~\cite{BKPO} as a limiting case,  modeling scanning electron microscopy of ultracold atomic gasses~\cite{Ott}. In such an experimental setting, a  Bose-Einstein condensate is affected by a narrow electronic beam, which in the meanfield approximation is described by a localized dissipative potential in the Gross-Pitaevskii equation~\cite{Pitaevskii}. Similar scenarios with the spatially confined absorption of nonlinear waves can be implemented in other experimental setups, like nonlinear optical and  magnon waveguides, plasmonic nanostructures, exciton-polariton condensates, etc. (see e.g.~\cite{ZezKon2016,ZKBO} for  schematics of possible systems).}

In what follows, we present our main results mainly for the perfectly absorbed flows supported by a $\delta$-function-shaped dissipation situated at $x=0$, and therefore we assume $\gamma>0$. In the meantime, most of our results can be generalized straightforwardly on the case of a lasing potential by inverting the sign of $\gamma$.  

Looking for stationary states $\Psi = e^{-i\mu t} \psi(x)$,  where real $\mu$ has the meaning of the chemical potential of the condensate, we use the hydrodynamic representation of the time-independent wavefunction $\psi(x)=\rho(x)\exp\left\{i\int_0^x v(s)dx\right\}$, where $|\rho| = |\psi|$ is the amplitude of the wavefunction, and $v(x)$ is the hydrodynamics velocity. The respective  current density, $j(x)$, is defined as  $j(x) = 2 v(x) |\psi|^2$.  Now Eq.~\eqref{eq:main}   reduces to the  system 
\begin{eqnarray}
\label{eq:rhoxx}
\rho_{xx} + \left(R^2 + \frac{J^2}{4R^4}\right)\rho - \rho^3 - \frac{j^2}{4\rho^3}=0,
\\[2mm]
\label{eq:jx}
j_x + 2\gamma\delta(x)\rho^2=0.
\end{eqnarray}

We are looking for perfectly absorbed flows directed from the left and the right infinities towards the center. The corresponding solutions  are determined by  the  boundary conditions
\begin{eqnarray}
\label{eq:as}
\lim_{x\to\pm \infty}\rho(x) = R, \quad \lim_{x\to\pm \infty}j(x) = \mp J,
\end{eqnarray}
where constants $R\geq 0$ and $J\geq 0$ set the  background amplitude  and the magnitude of the flow at the infinities. These boundary conditions fix the chemical potential 
\begin{equation}
\label{chim-pot}
\mu = R^2  + \frac{J^2}{4R^4}.
\end{equation}

Since the perfectly absorbed  flows are directed from the infinity to the center, the limiting (below also called background) current density    is negative ($-J$) for large positive $x$ and positive ($+J$) for large negative $x$. The case of lasing solutions emitted by a $\delta$-function amplifying potential {with $\gamma<0$} can be addressed by assuming that $J<0$.  {In this situation it follows   from (\ref{eq:as})    that the current $j(x)$ is positive for   $x>0$ and is negative for $x<0$.}

It follows from Eq.~(\ref{eq:jx})  that the current density is the step function $j(x) = -J\, \mathrm{sign}\, x$,  and the background current density is related to the amplitude at $x=0$ as  
\begin{align}
\label{J-rho}
J=\gamma \rho^2(0).
\end{align}
Using that $j^2(x) = J^2$ is constant for $x\neq0$, and integrating Eq.~(\ref{eq:rhoxx}) we obtain a first-order nonlinear differential equation, in which $J$ plays the role of a parameter:
\begin{equation}
\label{eq:int}
\rho_x^2 + \left(R^2 + \frac{J^2}{4R^4}\right)\rho^2 - \frac{\rho^4}{2} + \frac{J^2}{4\rho^2} = 
\frac{R^4}{2}+ \frac{J^2}{2R^2}.
\end{equation}
Using relation (\ref{J-rho}), from Eq.~\eqref{eq:int} one can express   the  derivative of the amplitude at the origin through the parameters of the problem:
\begin{equation}
\label{eq:rhox}
[\rho_x(0)]^2 = \frac{1}{4\gamma^2R^4}(\gamma R^2 - J)^2(2R^4 - \gamma J).
\end{equation}

\section{The main results}
\label{sec:results}

\subsection{Symmetric and asymmetric perfectly absorbed flows}
\label{sec:types}
Let us now discuss possible types of perfectly absorbed solutions that can be found in the introduced model.  First, we notice that there  exists an immune to the dissipation solution in the form of a dark soliton pinned to $x=0$~\cite{BKPO}: $\psi = R \tanh(R x/\sqrt{2})$.  
This solution does not correspond to a real physical absorber since the corresponding current is identically zero: $j(x)\equiv 0$.  {Solutions of the second type}   correspond to the  constant-amplitude nonlinear CPA modes and have uniform amplitudes  $\rho(x)=R$. Such modes are characterized by the background current densities $J=\gamma R^2$ and represent the direct nonlinear generalization, parametrized by $R$,  of the linear CPA  corresponding to the SS of the absorbing $\delta$-potential~\cite{Mostafazadeh2006}. These nonlinear modes exist  for arbitrarily  strong dissipation $\gamma$. 

\begin{figure}
	\begin{center}\par
		\includegraphics[width=0.99\columnwidth]{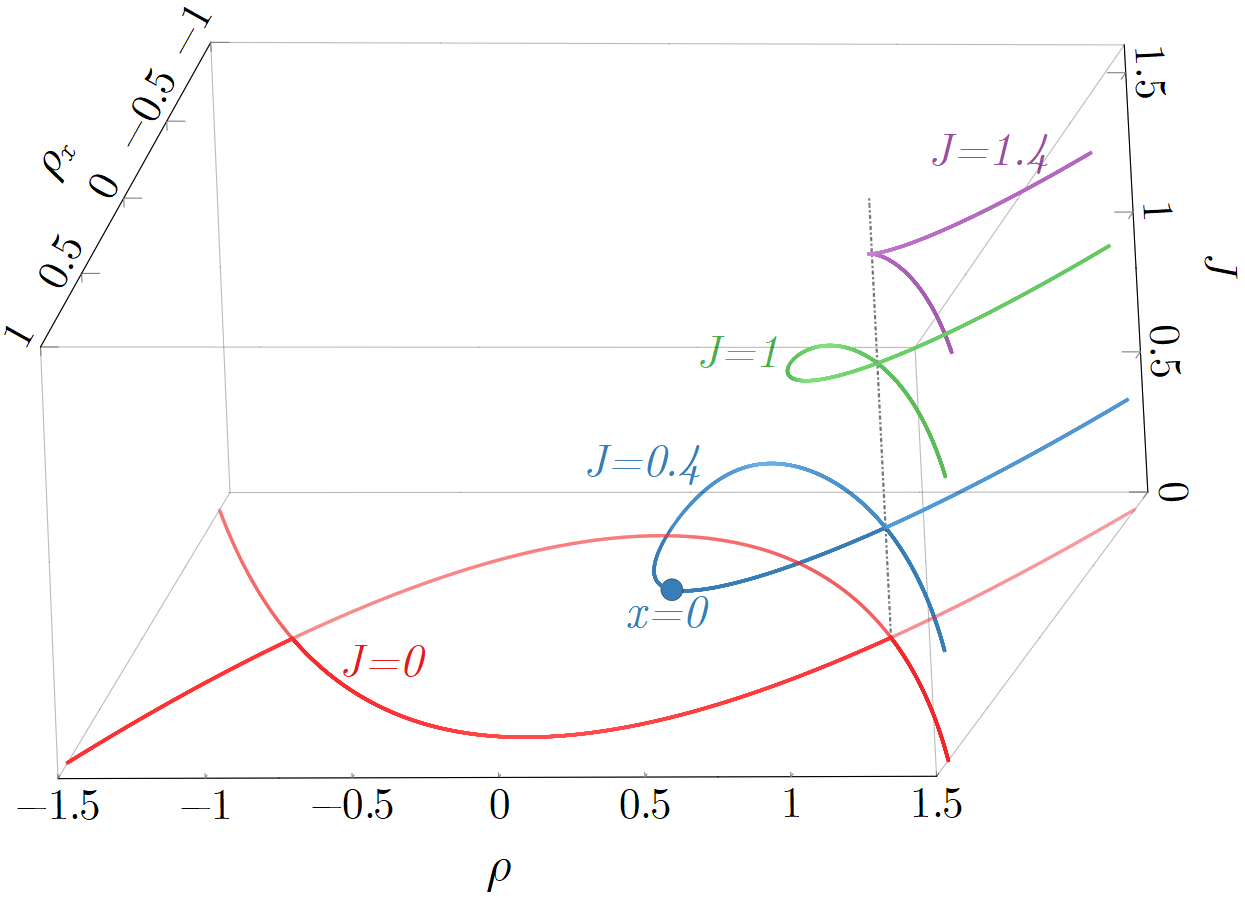}
	\end{center}
	\caption{Examples of orbits generated by the differential equation (\ref{eq:int}) in the phase space $(\rho, \rho_x)$ for several different values of $J$.  The heteroclinic orbits shown with red lines correspond to the dark soliton solution that is immune to the dissipation and bears zero background current $J=0$. Homoclinic orbits   with $J=0.4, 1, 1.4$ correspond to   perfectly absorbed flows with a density dip in a uniform background.  Each homoclinic orbit passes through the saddle point $(\rho, \rho_x)=(R,0)$, where $R$ is the background amplitude. Depending on the position of the point that corresponds to the dissipative spot, i.e., $x=0$ in our case, each orbit can represent either   a symmetric or an asymmetric perfectly absorbed flow. In this figure,  the background amplitude of nonlinear flows is fixed as $R=1$.}
	\label{fig:orbits}
\end{figure}

{Solutions of the third type are characterized by a spatially nonuniform amplitude with a dip.}  These solutions do not exist in the linear limit. They feature nonzero current density at the dissipative spot and have nontrivial and asymmetric amplitude landscapes. The possibility of   existence of such currents becomes evident from the inspection of the phase space $(\rho, \rho_x)$ corresponding to the identity (\ref{eq:int}) which for any $J$ in the interval $(0, \sqrt{2}R^3)$ features a homoclinic   orbit connecting the saddle point $(\rho, \rho_x)=(R,0)$ to itself (according to the introduced boundary conditions, the latter saddle point corresponds to $x=\pm \infty$).  Further analysis of the differential equation (\ref{eq:int}) shows that   the homoclinic orbits exemplified in Fig.~\ref{fig:orbits} are associated with exact solutions of the form
\begin{equation}
\rho^2(x) = R^2 - D^2\sech^2\left(\frac{ D(x-x_0)}{\sqrt{2}}\right),
\end{equation}
where the  new parameters $D$ and  $x_0$ are defined from the relations
\begin{eqnarray}
\label{eq:a^2}
D^2 &=& R^2 -   \frac{J^2}{2R^4},\\[3mm]
\label{eq:x0}
x_0 &=& \pm \frac{2R^2}{\sqrt{2R^6 - J^2}}\,\acosh\left(\frac{1}{R^2}\sqrt{\frac{2R^6 - J^2}{2(R^2-J/\gamma)}}\right).
\end{eqnarray}
Notice that the quantity $D^2$ which characterizes the depth  of the dip in the    squared amplitude is  positive provided that $J<\sqrt{2}R^3$. Additionally, for the solution to be meaningful, two more  constraints must be imposed. The first   condition reads 
\begin{equation}
\label{eq:cond1}
J<\gamma R^2.
\end{equation}
This condition is necessary to   guarantee that the argument of $\acosh$ in Eq.~\eqref{eq:x0} is real. 
In the asymptotic  limit where $J$ approaches $\gamma R^2$ from below,   the argument of $\acosh$  diverges, and the position of the intensity dip $x_0$ tends to $\pm \infty$. Another condition that needs to be imposed for the solutions to be meaningful  reads 
\begin{equation}
\label{eq:cond2}
J\leq 2  R^4/\gamma.
\end{equation}
It implies that the right-hand side of (\ref{eq:rhox}) is nonnegative, and the argument of $\acosh$ in Eq.~\eqref{eq:x0} is greater than or equal to unity.

\begin{figure}
	\begin{center}
		\includegraphics[width=0.99\columnwidth]{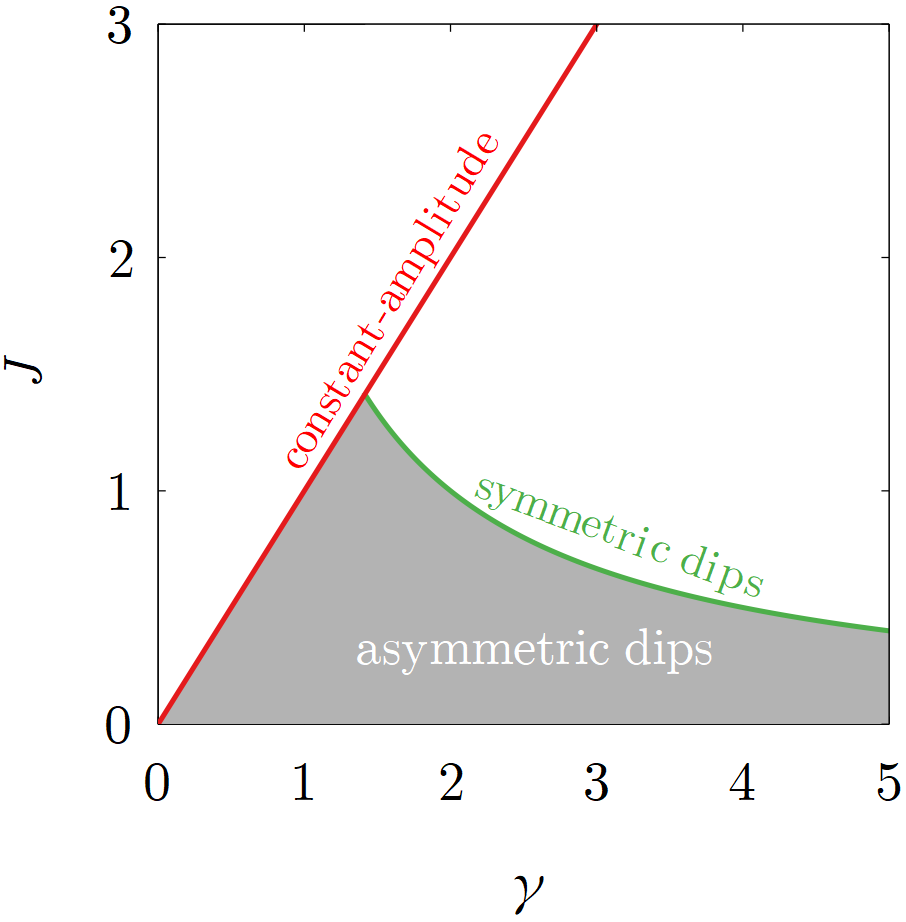}
	\end{center}
	\caption{The existence diagram {for solutions of three different types  discussed in  Sec.~\ref{sec:types}.} The diagram is shown in the plane $J$ {\em vs.} $\gamma$ for fixed background density $R=1$. The  gray domain corresponds to asymmetric dips, whereas red and green lines correspond to the constant-amplitude solutions and symmetric dips, respectively.}
	\label{fig:diag}
\end{figure}

\begin{figure}
	\begin{center}
		\includegraphics[width=0.99\columnwidth]{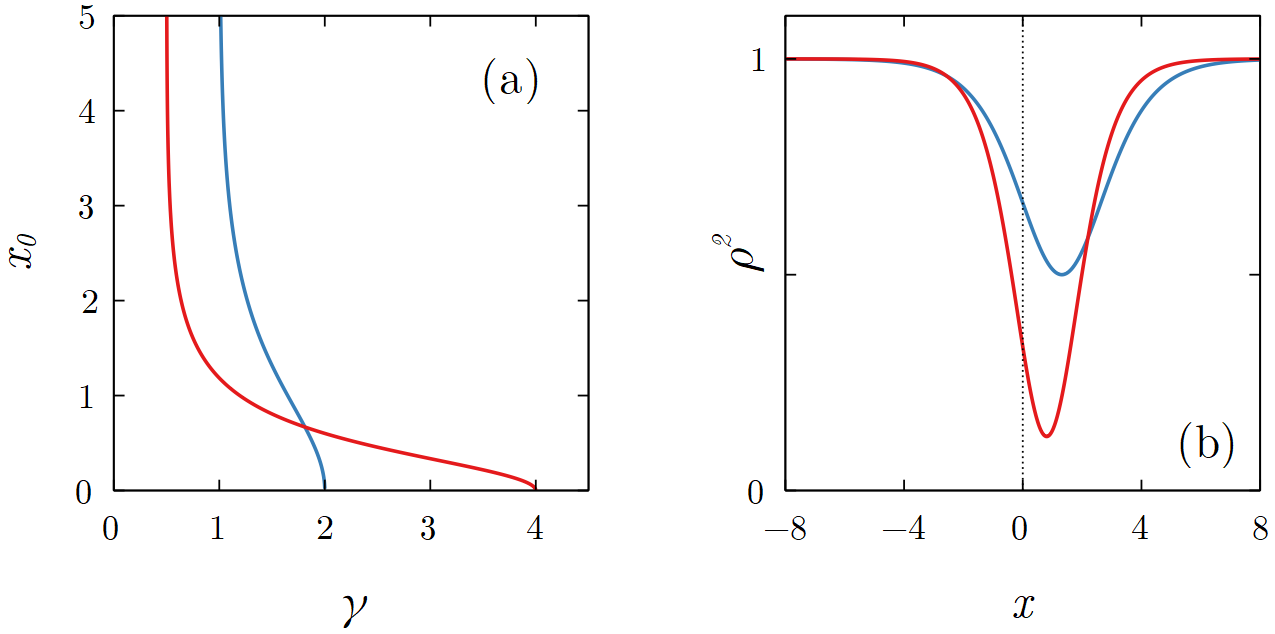}
	\end{center}
	\caption{(a) Dependencies of the position of the amplitude dip $x_0$ defined by (\ref{eq:x0}) on the absorption strength $\gamma$ for the fixed values of the background current: $J=1/2$  (red curve) and $J=1$ (blue curve). (b) Squared amplitude of asymmetric states that exist at $\gamma=3/2$ with $J=1/2$ (red curve) and $J=1$ (blue curve). In both panels $R=1$. Only solutions with $x_0\geq 0$ are shown; there also coexist their mirror counterparts with $x_0\leq 0$.}
	\label{fig:x0}
\end{figure}

Notice that found solutions are generically \emph{asymmetric}, i.e., the minimum of amplitude is achieved at $x_0$ which is generically different from zero.  Notice also that found states are \emph{essentially nonlinear}  { since}   they cannot be reduced to the linear limit by sending the nonlinearity coefficient to zero, i.e., they have no counterpart 
in the asymptotically linear limit of small background amplitudes      $R\to 0$. Such asymmetric perfectly absorbed states  { exist} in pairs, with positive and negative $x_0$,  which is reflected by the $\pm$ sign in (\ref{eq:x0}).  To summarize the situation,   in Fig.~\ref{fig:diag} we show a {representative  existence diagram for perfectly absorbed solutions of different types. Since the main  {characteristic} of a perfectly absorbed solution is the associated flux $J$, the existence diagram is presented  in the plane  $J$ vs. $\gamma$ for fixed background amplitude $R$}.   Emerging of asymmetric nonlinear states upon the increase of the dissipation strength is illustrated in Fig.~\ref{fig:x0}(a) for two fixed values of the background current $J$. Asymmetric states emerge when the dissipation strength exceeds   $J/R^2$ (the latter value corresponds to the dissipation that  {is} necessary to support the constant-amplitude CPA state with the given current $J$).  Emerging nonlinear states  are characterized by the   infinitely large position of the amplitude dip $x_0$ (mathematically diverging $x_0$ is explained by the fact that the argument of $\acosh$ in (\ref{eq:x0}) is infinitely large). The further increase of the dissipation strength $\gamma$ decreases $x_0$ and eventually, at $\gamma=2R^4/J$, the amplitude distribution becomes symmetric with $x_0=0$.  Thus symmetric perfectly absorbed states correspond  to the equality sign in  (\ref{eq:cond2}). The existence of symmetric dip solutions  can be explained by the fact that in a realistic system that cannot support infinitely large background currents $J$, one should expect that the  increase of the dissipation strength   destroys the constant amplitude of the steadily absorbed state and eventually results in the   \emph{decrease} of the background current rather than to its increase (compare red and green curves in Fig.~\ref{fig:diag}). This behavior can be attributed to the macroscopic Zeno effect studied previously in nonlinear waveguides with localized dissipation \cite{ZKBO}.

\subsection{Stability of perfectly absorbed nonlinear currents}

In order to study spectral stability of the perfectly absorbed flows, we consider a perturbed stationary mode in the form $\Psi(x,t) = e^{-i\mu t} [\psi(x) + u(x)e^{i\omega t} + v^*(x)e^{-i\omega^* t}]$, where $u$ and $v$ are small-amplitude perturbations. {A linearization of}    the main equation (\ref{eq:main}) with respect of $u$ and $v$  leads to the system
\begin{eqnarray}
\label{eq:ls1}
(\partial^2_x + \mu + i\gamma \delta(x) - 2|\psi|^2) u - \psi^2v&=&\omega u,\\[1mm]
\label{eq:ls2}
(\partial^2_x + \mu - i\gamma \delta(x) - 2|\psi|^2) v - (\psi^{2})^*u&=&-\omega v.
\end{eqnarray}
The spectrum of eigenvalues $\omega$ associated with    bounded eigenfunctions $u(x)$ and $v(x)$  characterizes stability of the stationary solution: the perfectly absorbed flow is unstable if there is an eigenvalue $\omega$ with negative imaginary   part: $\mathrm{Im}\, \omega <  0$. Since for the defocusing nonlinearity  the modulational instability of the uniform background is absent, and it is intuitively clear that the thin absorbing layer cannot excite spatially unbounded unstable modes, we expect that   the eventual instability can be caused only by spatially localized eigenmodes. We are therefore interested in eigenfunctions  $u$, $v$ that decay as $x\to \pm \infty$.

In general, eigenvalue problem (\ref{eq:ls1})--(\ref{eq:ls2}) can only be solved numerically. However, for the constant-amplitude solutions, the analytical treatment is possible. Let us first  consider this problem only in the     right half-axis $x>0$, where the stationary constant-amplitude wavefunction reads $\psi(x) = R e^{-i\gamma x/2}$, and chemical potential reads $\mu=R^2 + \gamma^2/4$. Using substitutions 
$u(x) = U(x) e^{-i\gamma x/2}$, $v(x) = V(x) e^{i\gamma x/2}$, the stability  equations take the form of a constant-coefficient ODE problem
\begin{equation}
\label{eq:ODE1}
\partial_x\bU + {\cal A} \bU=0,
\end{equation}
where 
\begin{eqnarray}
\label{eq:ODE2}
\bU &=& \left(\begin{array}{c} U\\ \partial_x U\\ V\\ \partial_x V\end{array} \right), \\[2mm]
\mbox{\quad and\quad } {\cal A}& =& \left(\begin{array}{cccc} 0& -1& 0&0\\
-\omega-R^2& - i\gamma& - R^2&0\\
0&0&0& - 1\\
- R^2&0& \omega - R^2&i\gamma
\end{array} \right).
\end{eqnarray}
Searching  for   solutions proportional to  $e^{-kx}$,  the exponent $k$ is determined from the characteristic equation for matrix ${\cal A}$:
\begin{equation}
\label{eq:poly}
k^4+(\gamma^2 - 2R^2) k^2 + 2 i\gamma   \omega  k -\omega^2=0.
\end{equation}
For solutions that decay at $x\to\infty$,   roots $k$ must belong to  the right complex half-plane. Using the  Routh-Hurwitz theorem \cite{Gantmacher},  we conclude that   for any $\omega$ with negative imaginary part the characteristic equation (\ref{eq:poly})   has exactly two  roots $k_{1,2}$  with positive real parts. Generically, 
those two roots are different, 
and the most general solution that decays at $x\to\infty$ is a linear combination of two independent exponents. In terms of functions $u(x)$ and $v(x)$ this solution    has the form
\begin{eqnarray}
u_+(x) &=& (U_{1,+} e^{-k_1 x} + U_{2,+} e^{-k_2 x})e^{-i\gamma x/2},\\
v_+(x) &=& (U_{1,+}\Lambda(k_1) e^{-k_1 x} + U_{2,+}\Lambda(k_2) e^{-k_2 x})e^{i\gamma x/2},
\end{eqnarray}
where $U_{1,2,+}$ are {complex-valued  constant  coefficients}, and  $\Lambda(k) = (k^2 + i\gamma k  - \omega)/ R^2  - 1$.

Next, we consider  eigenvalue problem (\ref{eq:ls1})--(\ref{eq:ls2}) only in the left half-axis $x<0$. The analysis can be performed in a similar by replacing   $\gamma$ to $-\gamma$ and $k$ with  $-k$ (because now we consider growing solutions $U,V\propto e^{k x}$). As a result, the general solution   reads
\begin{eqnarray}
u_-(x) &=& (U_{1,-} e^{k_1 x} + U_{2,-} e^{k_2 x})e^{i\gamma x/2},\\
v_-(x) &=& (U_{1,-}\Lambda(k_1) e^{\lambda_1 x} + U_{2,-}\Lambda(k_2) e^{k_2 x})e^{-i\gamma x/2}.
\end{eqnarray}

Applying the continuity conditions $u_+(0)=u_-(0)$, $v_+(0)=v_-(0)$  and using that $\Lambda(k_1) \ne \Lambda(k_2)$, we conclude that $U_{1,+} = U_{1,-} = U_1$ and $U_{2,+} = U_{2,-} = U_2$.  Integrating  equations (\ref{eq:ls1})--(\ref{eq:ls2})  across $x=0$, we obtain conditions for jumps of the derivatives:
\begin{eqnarray}
u_{+,x}(0) - u_{-,x}(0) &=& - i \gamma u_\pm(0),\\
v_{+,x}(0) - v_{-,x}(0) &=&  i \gamma v_\pm(0).
\end{eqnarray}
It is easy to check that these conditions imply $U_1=U_2=0$; therefore no unstable localized mode is possible. 

For the peculiar case when the characteristic  equation (\ref{eq:poly}) has a double root $k=k_0$ in the right half-plane, inspecting the structure of this polynomial one can establish that the latter root is positive: $k_0>0$, and, respectively, if the   corresponding eigenvalue $\omega_0$ exists, then it is   purely imaginary: $\omega_0 = i\lambda_0$, $\lambda_0<0$. In this case without loss of generality one can consider solutions of the form $U(x) = V^*(x)$. The solution has the form   $U(x) = (a_1  \pm  a_2x)e^{\mp k_0x}$, where upper and lower signs correspond to $x>0$ and $x<0$,  respectively, and $a_{1,2}$ are some coefficients.  From the condition of the derivative jump it follows that $a_1 = k a_2$. On the other hand, from the ODE system (\ref{eq:ODE1})-(\ref{eq:ODE2}) one can derive $(k_0^2+i\lambda_0+R^2)a_2+R^2a_2^*=0$, which is impossible for real $k_0$ and $\lambda_0$. Therefore, in the case of the double root the instability of constant-amplitude currents  cannot take place. 

Thus we have demonstrated that no localized eigenmodes of the stability problem  exists for any $\omega$ with nonzero real part. This implies stability of constant-amplitude nonlinear CPA states.

For dark solitons and dip solutions linear stability equations  (\ref{eq:ls1})--(\ref{eq:ls2}) does not admit simple analytical treatment, but the spectrum of eigenvalues $\omega$ can be computed numerically. A systematic stability study demonstrates that these perfectly absorbed solutions are also stable for all parameters in their existence domain.


\section{Conclusion}
\label{sec:concl}

In this paper, we have studied nonlinear stationary flows perfectly absorbed by an idealized infinitely narrow dissipative spot of infinite strength, which can be modeled by a Dirac $\delta$-function potential with a purely imaginary amplitude. The found solutions have been classified   {into} three types. Solutions of the first type have the form of conventional dark solitons. Their amplitude is identically zero at the dissipative spot, and these solutions are therefore immune to dissipation and do not generate superfluid flows. Solutions of the second class have constant amplitude and represent  {a} direct nonlinear generalization of linear coherently absorbed modes corresponding to the spectral singularities of the underlying $\delta$-function potential.  Solutions of the third type represent intensity dips in the uniform amplitude. Remarkably, these solutions do not have linear counterparts, and their amplitude distribution can be asymmetric.  Constant-amplitude CPA flows and (a)symmetric dips are supported by superfluid flows directed towards the dissipation. Using the linear stability approach, we have demonstrated analytically that constant-amplitude flows are stable.  {The} stability of dark solitons and dip solutions has been confirmed numerically.  

\begin{acknowledgments}
Work of D.A.Z. is funded by Russian Foundation for Basic
Research (RFBR) according to the research project No.~19-02-00193. V.V.K. acknowledges supported from the Portuguese Foundation for Science and Technology (FCT) under Contract no. UIDB/00618/2020.
\end{acknowledgments}

\end{document}